\documentclass[prd,superscriptaddress,eqsecnum,twocolumn,amsfonts,showpacs]{revtex4}
\usepackage{epsfig}

\newcommand{\be}{\begin{equation}}
\newcommand{\ee}{\end{equation}}
\newcommand{\bea}{\begin{eqnarray}}
\newcommand{\eea}{\end{eqnarray}}

\newcommand{\nn}{\nonumber}

\newcommand{\om}{\omega}

\begin{document}

\preprint{ \parbox{1.5in}{\leftline{hep-th/??????}}}

\title{Dynamical chiral symmetry breaking with  Minkowski space integral
 representations}

\author{V.~\v{S}auli}
\affiliation{Department of Theoretical Physics, Nuclear Physics
Institute, \v{R}e\v{z} near Prague, CZ-25068, Czech Republic}
\affiliation{CFTP and Departamento de F\'{\i}sica, Instituto Superior
T\'ecnico, Av. Rovisco Pais, 1049-001 Lisbon, Portugal }
\author{J.~Adam, Jr.}
\affiliation{Department of Theoretical Physics,
Nuclear Physics Institute, \v{R}e\v{z} near Prague, CZ-25068,
Czech Republic}
\author{P.~Bicudo}
\affiliation{CFTP and Departamento de F\'{\i}sica,
Instituto Superior T\'ecnico, Av. Rovisco Pais, 1049-001 Lisbon,
Portugal }

\begin{abstract}
The fermion propagator is studied in the whole Minkowski space with
the help of the Schwinger-Dyson equations. Various integral
representations are employed to get solutions for the dynamical breaking
of chiral symmetry in different regimes of the coupling constant.
In particular, in the case of massive boson,
we extend the singularity structure of the fermion  propagator to
the two real pole Ansatz. 
\end{abstract}

\pacs{11.10.St, 11.15.Tk}
\maketitle
%

%%%%%%%%%%%%%%%%%%%%%%%%%%%%%%%%%%%%%%%%%%%%%%%%%%%%%%%%%%%%%%%%%%%%%%%%%%%%%%%%%%%%%

In this brief report we investigate a possible scenario of 
 dynamical mass generation and estimate the timelike structure of the
fermion propagator.  This phenomenon, dubbed also as dynamical chiral symmetry breaking, 
requires intrinsically non-perturbative tools since the particle masses 
can be fully generated via loop contributions.
In the framework of Schwinger-Dyson equations (SDEs)  
 we explore the fermion mass function and the
fermion propagator in Minkowski space. We develop a novel integral
technique to solve with reasonable precision a SDE which has only
been addressed at one-loop order \cite{BICUDO} (for Yukawa model).
Here, we consider a gauge theory and resort to a simple
quenched approximation with the massive gauge boson transverse mode.
The effective coupling is then regulated by a Pauli-Villars cutoff
$\Lambda$.

The main result of this paper is to show that for the scaling $M/\Lambda<<1$ 
(walking Technicolor) the analytical structure of the exact propagator
is given the Lehmann representation with one real pole in this propagator.
Increasing the ratio $M/\Lambda$, we explicitly show that two pole Ansatz plus 
the corresponding generalized integral representation for the exact propagator is 
fully adequate for the description of dynamical chiral symmetry breaking in this phase. 
The novel integral representation which goes beyond the Lehmann representation is 
introduced for this purpose. Within the presented framework 
we achieve larger value of the scaling  $M/\Lambda\simeq 0.1$.

%%%%%%%%%%%%%%%%%%%%%%%%%%%%%%%%%%%%%%%%%%%%%%%%%%%%%%%%%%%%%%%%%%%%%%%%%%%%%
 In a parity conserving
theory the general form of the fermion propagator reads
\begin{equation}
S(p)=\frac{F(p^2)}{\not p-M(p^2)} \, .
\end{equation}
For simplicity we assume $F(p^2)=1$, which is reasonable
approximation for gauge theories in the Landau gauge. The SDE for the
mass function $M$ is modeled in the following manner
\bea 
&&M(p^2)=S^{-1}_{0}(p)- S^{-1}(p)=ig^2\int\frac{d^4q}{(2\pi)^4} \gamma^{\alpha}S(q)\gamma^{\beta}
\nn \\
&&
\left[-g^{\alpha\beta}+\frac{l_{\alpha}l_{\beta}}{l^2}\right]
\left[\frac{1}{l^2-m_B^2+i\epsilon}-\frac{1}{l^2-{\Lambda}^2+i\epsilon}\right] \, ,
\nn
\eea
where $l=p-q$, the constant $g^2$ implicitly absorbs a possible group prefactor, and $S_{0}(p)$
represents the free fermion propagator. 
In such approximation the effective coupling does not run
logarithmically, but it is constant up to a  scale $\Lambda$ 
where it rapidly vanishes (i.e. it runs with power behaviour).
Notice that in what concerns QCD, lowering the cutoff to the scale of
$\Lambda_{QCD}\simeq 250$ MeV and keeping the coupling large enough
such that constituent (infrared) quark  mass $M\simeq \Lambda$ can be
regarded as an approximation of QCD, while when $M<< \Lambda$, the limit of    walking
 Technicolors is  modeled  \cite{HOLDOM1985,YABAMA1986}, for a recent treatment within the SDEs framework see \cite{JAPONCI}.

A little is known about the full Minkowski solution of SDEs
in strong coupling field theories, hence we can refer here 
the paper of Fukuda and Kugo \cite{FUKKUG}
Furthermore, the timelike structure of Greens function 
as it is read from the Euclidean counterparts is not reliably known
\cite{FISCHER}. 
The main aim of this report  is to present   the direct solutions  in
Minkowski space, assuming  a spectral and a generalized integral
representation of the propagator for this purpose. 

 In order to carefully compare our Minkowski solutions
with the spacelike part obtained independently in Euclidean space,  we also 
perform the Wick rotation $iq_0\rightarrow q_4$ and solve SDE in Euclidean space
After the angular integration the Euclidean  SDE 
reads \cite{MANA1974},
\begin{eqnarray}
&&\label{euclid} M(p_E^2)=\frac{3\alpha}{8\pi }\,
\int_0^{\infty}d q^2_E \,\frac{M(q^2_E)\left[B(q,p,m_{B})-B(q,p,\Lambda)\right]}
{p_E^2 \left[q^2_E+M^2(q_E^2)\right]}
\nn \\
&&B(q,p,x)=-x^2+\sqrt{\lambda(-p^2_E,-q^2_E,x^2)}\, ,
\end{eqnarray}
where $q^2_E=-q^2$ (and similarly for $p$), $\alpha=g^2/(4\pi)$
and the symbol $\lambda$ stands for the triangle K\"allen function,
$\lambda(x,y,z)=(x-y-z)^2-4yz $ .

The solution of the Eq. ~(\ref{euclid}) is well known:
 for the coupling below certain
critical value $\alpha_c$ there exists only a trivial solution
$M(p_E^2)=0$, while for $\alpha > \alpha_c$ we get a non-trivial
mass function. The value of
the critical coupling $\alpha_c$ depends on the details of the
kernel, especially on the  finite ratio $R=m_B/\Lambda$, noting that
 for $m_B<<\Lambda$  the critical
coupling constant  $\alpha_c\simeq \pi/3$ coincides with the one
obtained in the ladder approximation for the electron propagator in
the strong coupling QED, where the well-known exponential Miransky scaling
is exhibited:
\be
M(0)\simeq const\cdot \Lambda e^{-\left[\pi(\alpha/\alpha_c-1)^{-1/2}\right]} \, .
\ee

In the first part of our SDE Minkowski study we assume spectral representation 
with a single real pole in
the propagator and derive the Unitary Equations in their full
non-linearized form. The solutions of Schwinger-Dyson equations obtained by
the spectral method  has been already  calculated  for several models
\cite{LACO,SAULIJHEP}. Stressed that  in any case, the resulting spacelike 
parts of Greens functions under consideration \cite{LACO,SAULIJHEP},
were in  a good agreement with the solutions based on the Euclidean
formalism.

Assumed Lehmann spectral representation reads,
\begin{equation}
\label{sesmek}
S(p)=\int_{R^{+}} d x\, \frac{\not p\sigma_v(x)
+\sigma_s(x)}{p^2-x+i\epsilon}\, ,
\end{equation}
where $S$ has a pole at $p^2=m^2$, i.e.,
$\sigma_s(x)=rm\delta(x-m^2)+\sigma_s^c(x)$, where
 $r$ represents the  residuum. The function  $\sigma_s^c(x)$ is a  continuous
part  of the spectral  function starting to be non-zero from the
first branch point. Substituting the integral representation
(\ref{sesmek}) into our gap equation  written in
Minkowski space,
\be
\label{SDEB}
M(p^2)=i3g^2\int dx \int\frac{d^4q}{(2\pi)^4}G(p-q)
\frac{\sigma_s(x)}{q^2-x+i\epsilon}\, ,
\ee
one arrives to the following dispersion relation for the mass
function $M$,
\begin{eqnarray}
\label{pispunta}
&&M(p^2)= \int d\omega \frac{\rho_s(\omega)}{p^2-\omega+i\epsilon}
\, ,  \\
&&\rho_s(\omega)=\frac{\alpha}{(4\pi)}\int d x \sigma_s(x)
\left[X_0(\omega;m_B^2,x)- X_0(\omega;\Lambda^2,x)\right]\, ,
\nn \\
&&X_0(\omega;a,b)=\frac{\sqrt{\lambda(\omega,a,b)}}{\omega}
\Theta\left(\omega-(\sqrt{a}-\sqrt{b})^2\right)\, .
\nn
\end{eqnarray}

The imaginary part of the propagator
and the imaginary part of the dynamical mass function are simply related, 
this relation closes the system of the equations (\ref{pispunta}) employed. In our
approximation $Z=1$ and it is sufficient to consider the $S_s$ part
of the propagator,
\bea
\label{megan}
 S_s(p^2)&=&\frac{M(p^2)}{p^2-M^2(p^2)}
\nn \\
&=&
\Sigma_R
\frac{p^2-\Sigma_R^2-\Sigma_I^2}
{\left(p^2-\Sigma_R^2+\Sigma_I^2\right)^2+4\Sigma_R^2\Sigma_I^2}
\nn \\
&+&i\, \Sigma_I \frac{p^2+\Sigma_R^2+\Sigma_I^2}
{\left(p^2-\Sigma_R^2+\Sigma_I^2\right)^2+4\Sigma_R^2\Sigma_I^2}
\eea
where we  use a shorthand notation,
$\Sigma_R={\rm Re}\ M(p^2)\, ;
\Sigma_I={\rm Im}\ M(p^2)=-\pi\rho(p^2)$.
Comparing imaginary part of (\ref{megan}) with the imaginary part of the propagator 
${\rm Im}\, S_s(p^2)=-\pi\sigma_s^c(p^2)$,
we immediately get
\be
\sigma_s^c(p^2)=\rho_s(p^2)\frac{p^2+\Sigma_R^2+\Sigma_I^2}
{\left(p^2-\Sigma_R^2+\Sigma_I^2\right)^2+4\Sigma_R^2\Sigma_I^2} \, ,
\ee
which is nonzero  for  time-like momenta above the threshold. The
derivation of more general ``Unitary equations'' which takes into
account the  wave function renormalization is straightforward  (see
for instance \cite{LACO}).

The dispersive  (real) part of the mass functions is given by the
principal value integral
\be \Sigma_R(p^2)=  P\cdot \int d\omega
\frac{\rho_s(\omega)}{p^2-\omega} \, .
\ee
The principal value can be avoided by  using $\rho$ as given in
(\ref{pispunta}), which yields an ordinary regular integral over the
new kernel,
\be
\Sigma_R(p^2)=\frac{3\alpha}{4\pi} \int d x \sigma_s(x)
\left[J(p^2,x,m_B^2)-J(p^2,x,\Lambda^2)\right] \, , \nn
\ee
where $J$ results from the principal value integration of the
dispersion relation for $M$,
\begin{eqnarray}
&&J(p^2,x,z)= -\frac{\Theta(-\lambda_p)\sqrt{-\lambda_p}}{p^2}
\left[\frac{\pi}{2}+
{\rm arctg}\frac{p^2-x-z}{\sqrt{-\lambda_p}}\right]
\nn \\
&&-\frac{\Theta(\lambda_p)\sqrt(\lambda_p)}{p^2}
\ln\left|\frac{p^2-x-z+\frac{\lambda_p}{(\sqrt{x}+\sqrt{z})^2-p^2}}{p^2-x-z+\sqrt{\lambda_p}}\right|+
\nn \\
&&\frac{\Theta(\lambda_0)\sqrt{\lambda_0}}{p^2}
\ln\left|\frac{-x-z+\frac{\lambda_0}{(\sqrt{x}+\sqrt{z})^2}}{-x-z+\sqrt{\lambda_0}}\right|
+\frac{1}{2}\ln(16xz) \nn \, ,
\end{eqnarray}
where we have shortly written $\lambda_{p,(0)}=\lambda(p^2(0),x,z)$.

The residuum $r$ and the pole mass function $m$ are obtained
evaluating the dispersion relation and its derivative. The coupled
set of the integral equations above has been solved numerically by
iterations.

%Fig 4
\begin{figure}\vspace{-1cm}
{\epsfig{figure=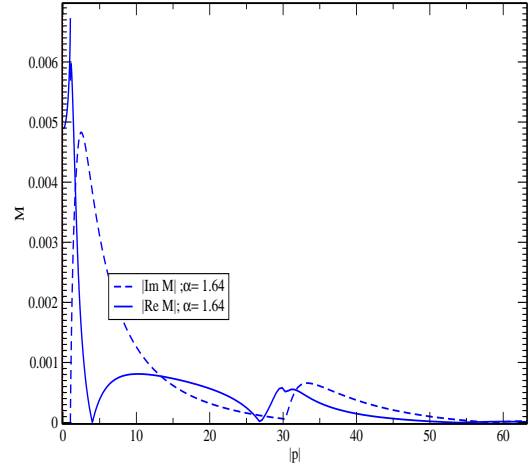,width=8truecm,height=8truecm,angle=270}}
\caption[caption]{ Absolute value of the mass function
at time-like momenta with single pole in the propagator. }
\label{time-like}
\end{figure}

\begin{figure}
{\epsfig{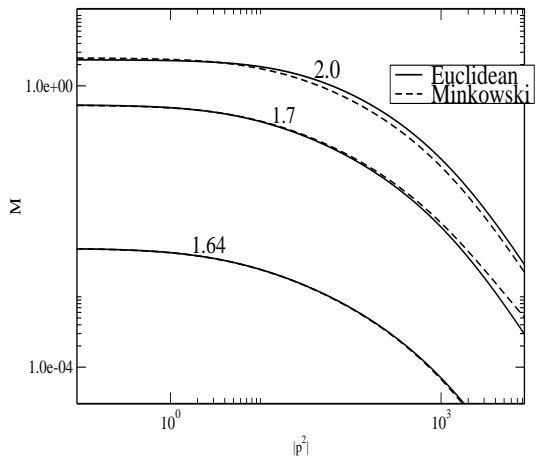}}
\caption[caption]{Dynamical mass function for space-like momenta for 
various coupling $\alpha=1.64; \, 1.7,\, 2.0$. 
Solid (dashed) lines stand for Euclidean (Minkowski) solutions.} \label{space-like}
\end{figure}

The Unitary equations provide solutions for time-like momenta above
the branch point $p^2>(m_B+m)^2$, below which the propagator is real.
The results on the negative axis of $p^2$ are easily obtained by a
regular integration either from (\ref{pispunta}) or from the
dispersion relation for $M$ (\ref{sesmek}).
The resulting timelike solution is presented in  Fig.~\ref{time-like}).
The results  presented here are calculated with
${\Lambda}/{m_B}=30$.

and we use the mass $m_B$ as a scale for all dimensionfull
quantities. The comparison of spectral Minkowski and Euclidean
solutions is shown in Fig.~\ref{space-like}. Thus, solving the Unitary
Equations and comparing the Minkowski solution to the Euclidean one,
we find rather nice agreement near the critical coupling. 
However, when the coupling becomes larger (say when $\alpha$ exceeds $\alpha_c$,
about  $\alpha$ 10 \%) a discrepancy
appears, since the employed spectral representation for the
propagator, with just  one pole, is no longer valid.
Retrospectively, the previous one loop analytical calculations
\cite{BICUDO} already found evidence for a more complex structure in
the propagator. Apparently  new singularities appear
in the propagator  for $\alpha>\alpha_T \, $ 
The coupling $\alpha_T$ was determined to be
$\, \alpha_T =1.73\pm0.02$ in our case. .

In what follows we continue with our study of the SDE in Minkowski
space relaxing our assumption on the spectral representation of the
fermion propagator. 
%
%Fig 2

\begin{figure}
\centerline{\epsfig{figure=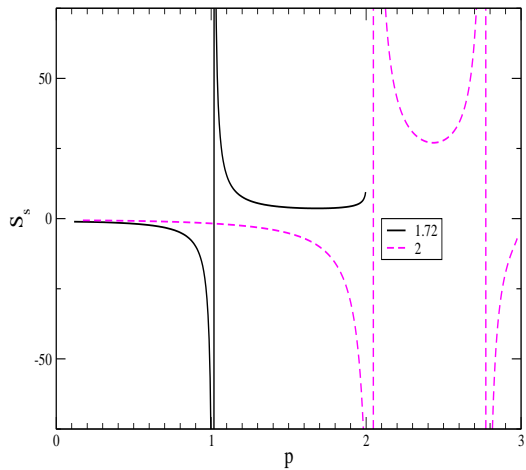,width=8truecm,height=8truecm,angle=270}}
\caption[caption]{(Color online)  Propagator singularities bellow threshold shown for two different coupling constant. The dashed (solid) line
stands for the case when the propagator function $S_s$ exhibits two (one) real poles below
the branch points.   \label{intuitive2}}
\end{figure}

The reason for the fail of one pole Ansatz is easy to understand: the dynamical mass is
an increasing function in the regime from $p^2=0$ to  the branch
point $p_B^2=(m+m_B)^2$. When the coupling is  large enough,
then the large enhancement below the branch point  maintains and the
dynamical mass function necessarily crosses the line $\sqrt{p^2}$.
Thus, for a certain set of parameters the propagator develops two
real poles which should be taken into account in the integral Ansatz.
The second real pole first appears at the branch point
$p_B^2=(m+m_B)^2$ and moves down towards the first one as the
coupling increases. To get a better view we draw this scenario in
in Fig.~\ref{intuitive2}. The lines displayed represent our 
numerical findings.

 The new integral Ansatz, which is consistent with the
solution of SDE in the regime where the two poles are present, reads
\bea \label{pair}
S_s(p^2)&=&\frac{m_-}{p^2-m^2_++i0}-\frac{m_-}{p^2-m^2_-+i0}
\nn \\
&+&\frac{1}{p^2-m^2_-+i0}\int d\omega \frac{\sigma_s^-(\omega)}{p^2-\omega+i0} \ .
\eea
The alternative equivalent representation can be obtained by
replacements $m_- \rightarrow - m_+, \sigma_s^- \rightarrow
\sigma_s^- $.

This Ansatz immediately implies the dispersive relation for the
dynamical mass function. Taking $F=1$ approximation again and substituting the formula (\ref{pair})
into the SDE  it leads after the integration over the
momenta to the following result for the function $\rho_s(p^2)$
\bea  \label{impart} &-& \rho_s(\om)\frac{(4\pi)^2}{g^2}=
m_-\left[X_0(\om;m_+^2,m_B^2)-X_0(\om;m_+^2,\Lambda^2)
\right. 
\nn \\
&&-    \left.X_0(\om;m_-^2,m_B^2)+X_0(\om;m_-^2,\Lambda^2)\right]
\nn \\
&+&\int dx\, \frac{\sigma_s^-(x)}{m_-^2-x}
\left[X_0(\omega;m_-^2,m_B^2)- X_0(\omega;{\Lambda}^2,m_-^2)\right]
\nn \\
&&-\int d x\frac{\sigma_s^-(x)}{m_-^2-x}
 \left[X_0(\omega;m_B^2,x)- X_0(\omega;{\Lambda}^2,x)\right]\, ,
\eea
where the first  line in (\ref{impart}) follows from the first two
terms in (\ref{pair}) and the second  line follows from the third
term in (\ref{pair}). 
The derivation of (\ref{impart}) is straightforward and follows the
same lines as in the case of the standard Lehmann representation. The
integration over the momentum is finite and it remains finite even
when $\Lambda$ is sent to infinity, which is a consequence of the
momentum behavior of our Ansatz (\ref{pair}).

The Unitary equations are modified since the integral
representation has changed. To derive 
 them let us  compare the imaginary and the real parts of the
propagator above the branch point $(m_B+m_+)^2$ (by definition,
$m_->m_+$).

On one side we get from the imaginary part of (\ref{pair})
\be
{\rm Im}\, S_s(p^2)=-\frac{\pi\sigma_s^{-}(p^2)}{p^2-m_-^2}
\ee
and the imaginary part of this function, computed with the SDE, is
still given by Eq.~(\ref{megan}). This implies that $S_s$ is real up
to the $p^2=(m_-+m_B)^2$.

Thus we get for the time-like momenta, such that $p^2>(m_-+m_B)^2$,
the following equation,
\be \label{due}
\sigma_s^-(p^2)=\rho(p^2)\frac{(p^2-m_-^2)(p^2+\Sigma_R^2+\Sigma_I^2)}
{\left(p^2-\Sigma_R^2+\Sigma_I^2\right)^2+4\Sigma_R^2\Sigma_I^2} \, .
\ee
This means that the momentum space Schwinger-Dyson equation turns
into two coupled regular equations (\ref{impart}) and (\ref{due})
relating the absorptive and dispersive parts of the propagator and
its inverse.
%Fig 3

The pole masses are necessarily expected below the threshold and
they are determined by the zeroes of the inverse of the propagator
i.e., $m_{\pm}=M(m_{\pm}^2) \,$ .
The real part of the mass $\Sigma_R$ entering Eq.~(\ref{due}) is
given by the principal value integration over $\rho$ in Eq.~(\ref{impart}) . 
 This leads to the following compact regular integral
equation for $\Sigma_R$,
\begin{eqnarray}
&-&\Sigma_R(p^2)\frac{(4\pi)^2}{g^2}=
m_-\left[J(p^2,m_-^2,m_B^2)-J(p^2,m_-^2,\Lambda^2)
\right.
\nn \\
&&\left.-J(p^2,m_+^2,m_B^2)+J(p^2,m_+^2,\Lambda^2)\right]
\nn \\
&+& \int d x\,
\frac{\sigma_s^-(x)}{m_-^2-x}
\left[
J(p^2,m_-^2,m_B^2)-J(p^2,m_-^2,{\Lambda}^2)\right.
\nn \\
&&-\left.J(p^2,x,m_B^2)+J(p^2,x,\Lambda^2)\right]\, .
\end{eqnarray}

The second pole appears for couplings stronger than $\alpha_{T} \simeq
1.73$.  In the
interval of the  couplings $ 1.73 < \alpha < 2.0$ the two-pole
representation of the propagator leads to solutions which, for
space-like momenta, agree rather well with the Euclidean ones.
% Fig 8
\begin{figure}
{\epsfig{figure=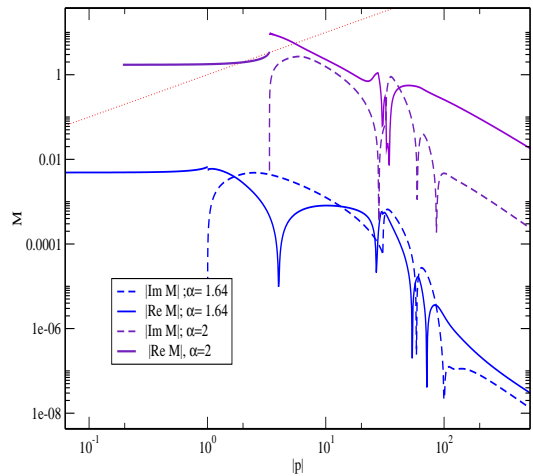,width=8truecm,height=8truecm,angle=270}}
\caption[caption]{(Color online) Absolute value of the mass function
at time-like regime of the momenta  in
log-log plot. The upper two lines represent the real and imaginary
parts of the mass function $M$ for $\alpha=2$, the two lines below
coincide with the data plotted in Fig. 2.} \label{time-like2}
\end{figure}
Our solution for the coupling $\alpha=2$ is added in the
Fig.~\ref{space-like} for spacelike and it is displayed in the Fig. ~\ref{time-like2} for timelike momenta.
Our  Minkowski solution  becomes unstable for $\alpha>2$ and starts to disagree with 
the spacelike Euclidean results for $\alpha>2$. To dive more deep into the chiral breaking phase
and thus achieve enhancement of the infrared mass would 
require a new reanalysis of up to now unknown, possibly complex, propagator singularities.

To conclude, the one-pole case is adequate to solve dynamical
chiral symmetry breaking near to the critical coupling, 
while the two pole fit is adequate to work with moderate
couplings, well into the phase transition region where the standard spectral 
representation deviates from the correct solution. For the later case,
a suitable integral representation of the propagator has been proposed.
\acknowledgements
V.~\v S and J.A. were supported by the grant GA CR 202/06/0746.

%%%%%%%%%%%%%%%%%%%%%%%%%%%%%%%%%%%%%%%%%%%%%%%%%%%%%%%%%%%%%%%%%%%%%%%


\begin{thebibliography}{99}
%
\bibitem{BICUDO}
P.~Bicudo, Phys. Rev. D{\bf69}, 074003 (2004).
%
%\bibitem{FASUS1981}
%E.~Farhi and L.~Susskind, Phys.~Rep.~{\bf 74}, 277 (1981).
%
%\bibitem{DISU1979}
%S.~Dimopoulos and L.~Susskind, Nucl.~Phys.~B{\bf 155}, 237 (1979).
%
%\bibitem{EILA1980}
%E.~Eichtem and K.~Lane, Phys.~Lett.~B{\bf 90}, 125  (1980).
%
%\bibitem{HILL1995}
%C.T.~Hill, Phys.~Lett.~B{\bf345}, 483 (1995).
%
%\bibitem{LANE1998}
%K.~Lane, Phys.~Lett.~B{\bf 433}, 96 (1998).
%
%\bibitem{GHTY2002}
%V.~Gusynin, M.~Hashimoto, M.~Tanabashi and K.~Yamawaki,
%Phys.~Rev.~{\bf D65}, 116008 (2002).
%
\bibitem{HOLDOM1985}
B.~Holdom, Phys.~Lett.~B{\bf 150}, 301 (1985).
%
\bibitem{YABAMA1986}
K.~Yamawaki, M.~Bando, and K.~Matumato,
Phys.~Rev.~Lett.~{\bf 56}, 1335 (1986).
%
%\bibitem{APWI1987}
%T.~Appelquist and L.C.R.~Wijewardhana,
%Phys.~Rev.~D{\bf 35}, 774 (1987); Phys.~Rev.~D{\bf 36}, 568 (1987).
%
\bibitem{JAPONCI}
M.~Harada, M.~Kurachi, K.~Yamawaki,Phys.~Rev.~D{\bf 68}, 076001 (2003);
M.~Kurachi and R.~Shrock, JHEP {\bf 0612}, 034 (2006).
%
%\bibitem{APRATEWI1998}
%T.~Appelquist, A.~Ratnaweera, J.~Terning, and L.C.R.~Wijewardhana,
%Phys.~Rev.~D{\bf 58}, 105017 (1998).
%
%\bibitem{NEPA2005}
%A.V.~Nesterenko, J.~Papavassiliou,
%Phys.~Rev.~D{\bf 71}, 016009 (2005).
%
%\bibitem{NEPA2005ii}
%A.V.~Nesterenko, J.~Papavassiliou,
%Int.~J.~Mod.~Phys.~A{\bf20}, 4622 (2005).
%
%\bibitem{BAMIST2005}
%A.P.~Bakulev, S.V.~Mikhailov, N.G.~Stefanis,
%Phys.~Rev.~D{\bf72}, 074014 (2005);
%Erratum-ibid. D{\bf 72}, 119908 (2005).
%
%
%
%\bibitem{HAWIRO1996}
%F.T.~Hawes, A.G.~Williams, C.D.~Roberts,
%Phys.~Rev.~D{\bf 54}, 5361 (1996).
%
\bibitem{MANA1974}
T.~Maskawa and H.~Nakajima,
Prog.~Theor.~Phys.~{\bf 52}, 1326 (1974).
%
%
%\bibitem{MONMUN}
%I.~Montvay and G.~Munster, "Quantum Fields On A Lattice",
%Cambridge Univ. Press. (1994), Cambridge, UK.
%
%\bibitem{HOROKE1992}
%L.C.~Hollenberg, C.D.~Roberts and B.H.~McKellar,
%Phys.~Rev.~C{\bf 46}, 2057 (1992).
%
%\bibitem{BUFRMI1997}
%M.~Burkardt, M.R.~Frank and K.L.~Mitchell,
%Phys.~Rev.~Lett.~{\bf 78}, 3059 (1997).
%
%\bibitem{ALDEFIMA2004}
%R.~Alkofer, W.~Detmold, C.S.~Fischer, P.~Maris,
%Phys.~Rev.~D{\bf 70}, 014014 (2004).
%
\bibitem{FUKKUG}
R.~Fukuda, T.~Kugo, Nucl.~Phys.~{B 117}, 250 (1976).
%
\bibitem{FISCHER}
R.~Alkofer, W.~Detmold, C.S.~Fischer, P.~Maris,
Phys.~Rev.~D{\bf 70}, 014014 (2004).
%
\bibitem{LACO}
V.~\v{S}auli, {\it Few Body Systems} {\bf 39}, 1-2, hep-ph/0412188.
%
\bibitem{SAULIJHEP}
V.~\v{S}auli,  {\it JHEP} 0302, 001 (2003).
%
%
%
%
%\appendix
\end{thebibliography}
\end{document}